\documentstyle[prb,aps,eqsecnum,multicol,epsfig]{revtex}

\newcommand{\bleq}{\ifpreprintsty
                   \else
                   \end{multicols}\vspace*{-3.5ex}{\tiny
                   \noindent\begin{tabular}[t]{c|}
                   \parbox{0.493\hsize}{~} \\ \hline \end{tabular}}
                   \fi} 
\newcommand{\eleq}{\ifpreprintsty
                  \else
                   {\tiny\hspace*{\fill}\begin{tabular}[t]{|c}\hline
                    \parbox{0.49\hsize}{~} \\
                    \end{tabular}}\vspace*{-2.5ex}\begin{multicols}{2}
                    \fi}
\newcommand{\bcols}{\ifpreprintsty\else\begin{multicols}{2}\fi}
\newcommand{\ecols}{\ifpreprintsty\else\end{multicols}\fi}

\begin{document}
\draft

\title{Spin fluctuations and pseudogap in the two-dimensional
  half-filled Hubbard model at weak coupling } 
\author{N. Dupuis }
\address{ Laboratoire de Physique des Solides, Associ\'e au CNRS, \\
  Universit\'e Paris-Sud, 91405 Orsay, France }
\date{March 27, 2002}
\maketitle

\begin{abstract} 
Starting from the Hubbard model in the weak-coupling limit, we derive
a spin-fermion model where the collective spin excitations are
described by a non-linear sigma 
model. This result is used to compute the fermion spectral function
$A({\bf k},\omega)$ in the low-temperature regime where the
antiferromagnetic (AF) coherence length is exponentially large
(``renormalized classical'' regime).  At the Fermi level, $A({\bf
k}_F,\omega)$ exhibits two peaks around $\pm\Delta_0$ (with
$\Delta_0$ the mean-field gap), which are precursors of the
zero-temperature AF bands, separated by a pseudogap.  
\end{abstract}

\pacs{PACS Numbers: 71.10.Fd, 75.10.Lp}

\bcols 


\section{Introduction}

In the last two decades, the discovery of heavy-fermion compounds,
high-$T_c$ superconductors and organic conductors has revived interest
in strongly correlated electron systems. Of particular interest are
metallic phases which, although conducting, are not described by
Landau's Fermi liquid theory because of the absence of well-defined
quasi-particle excitations. A well-known example is given by the
normal phase of high-$T_c$ 
superconductors. Instead of quasi-particles, these systems exhibit a
pseudogap at low energy as shown by many experiments.
\cite{Timusk99} Although the  
origin of the pseudogap is still under debate, it is generally
believed that antiferromagnetic (AF) fluctuations play a crucial role. 

In this paper, we consider the pseudogap issue on the basis of the
half-filled 2D Hubbard model. We consider only the weak-coupling limit
$U\ll t$ ($U$ is the local Coulomb repulsion and $t$ the intersite
hopping amplitude). [In the strong-coupling limit at half-filling, the
finite temperature paramagnetic phase is a Mott-Hubbard insulator with
a (charge) gap of order $U$. At $T=0$, there is a transition to a N\'eel
antiferromagnetic state.\cite{note1}] Although the ground-state is
AF, long-range order is destroyed by classical
fluctuations at any finite temperature, in agreement with the
Mermin-Wagner theorem. Nevertheless, below a crossover temperature
$T_X$ (of the order of the mean-field transition temperature), the
system enters a renormalized classical regime where AF correlations
start to grow exponentially. Contrary to the 3D case, at the
zero-temperature 2D phase transition the system goes directly into the
(N\'eel) ordered state where the fermion spectral function $A({\bf
  k},\omega)$ exhibits two well-defined 
quasi-particle (QP) peaks corresponding to the Bogoliubov QP's. 
By continuity, the two-peak structure in  $A({\bf k},\omega)$  cannot
disappear as soon as we raise the temperature. As pointed out in
Ref.~\onlinecite{Vilk97}, the only possible scenario is that at finite
but low temperature the fermion spectral function exhibits two
(broadened) peaks which are precursors of the $T=0$ Bogoliubov QP's,
separated by a pseudogap. We therefore expect the presence of
a pseudogap at finite temperature, due to the strong (classical) AF
fluctuations.

Clearly, traditional mean-field techniques fail to describe these
phenomena. For instance, the random-phase approximation (RPA) predicts a
finite temperature phase transition which is forbidden in 2D by the
Mermin-Wagner theorem. More sophisticated approaches are therefore
required. In the weak-coupling limit, the pseudogap formation has been
considered within the fluctuation 
exchange (FLEX) approximation \cite{Bickers89} and the
two-particle-self-consistent (TPSC) theory \cite{Vilk97,Vilk96,Vilk97a} which
both satisfy the Mermin-Wagner theorem. Only the TPSC theory predicts
the formation of a pseudogap in the fermion spectral function $A({\bf
k},\omega)$ at low temperature.

The aim of this paper is to describe an alternative approach to the
2D half-filled Hubbard model in the weak-coupling limit. We first
derive a spin-fermion model 
where the collective spin excitations are described by a non-linear
sigma model (NL$\sigma$M). The spin-wave velocity and the coupling
constant of the NL$\sigma$M are expressed in terms of the ground-state
properties of the system. Solving the NL$\sigma$M in a ``large-$\cal
N$'' limit, we then compute the fermion spectral function $A({\bf
  k},\omega)$ to lowest order in the spin-fermion interaction. At 
the Fermi level, $A({\bf k}_F,\omega)$ exhibits two peaks around
$\pm\Delta_0$ (with $\Delta_0$ the mean-field gap) which are
precursors of the zero-temperature AF bands, separated by a
pseudogap. We compare our results with those of the TPSC theory.  

\section{Model} 
The two-dimensional Hubbard model is defined by the Hamiltonian
\begin{equation}
H = -t \sum_{\langle {\bf r},{\bf r}' \rangle,\sigma} (
c^\dagger_{{\bf r}\sigma} c_{{\bf r}'\sigma} + {\rm h.c.}) 
+U \sum_{\bf r} n_{{\bf r}\uparrow} n_{{\bf r}\downarrow} ,
\label{Ham}
\end{equation}
where $t$ is the intersite hopping amplitude and $U$ the on-site
Coulomb repulsion. $  c_{{\bf r}\sigma}$ is a fermionic operator for a
$\sigma$-spin particle at site ${\bf r}$
($\sigma=\uparrow,\downarrow$), and $  n_{{\bf r}\sigma}= 
c^\dagger_{{\bf r}\sigma}  c_{{\bf r}\sigma}$.  
$\langle {\bf r},{\bf r}' \rangle$ denotes nearest-neighbor sites. 
We take the lattice spacing equal to unity and $\hbar=k_B=1$
throughout the paper.  

Since spin fluctuations play a crucial role in the Hubbard model at
half-filling, it is convenient to introduce auxiliary fields
describing these collective excitations. The standard approach is to
write the interaction part of the Hamiltonian in terms of charge and
spin fluctuations, i.e. $  n_{{\bf r}\uparrow}   n_{{\bf
r}\downarrow} = [(  c^\dagger_{\bf r}  c_{\bf r})^2 - ( 
c^\dagger_{\bf r}\sigma_z   c_{\bf r})^2]/4$, and then perform a
Hubbard-Stratonovich transformation by means of two (real) auxiliary
fields $\Delta_c$ and $\Delta_s$ [$c_{\bf r}=(c_{{\bf
r}\uparrow},c_{{\bf r}\downarrow})^T$]. Although this procedure
recovers the standard  
mean-field (or Hartree-Fock) theory of the N\'eel state within a
saddle-point approximation, it leads to a loss of spin-rotation
invariance and does not allow to obtain the spin-wave
excitations. Alternatively, one could write $n_{{\bf r}\uparrow}
n_{{\bf r}\downarrow}$ in a spin-rotation invariant form,
e.g. $n_{{\bf r}\uparrow} n_{{\bf r}\downarrow}=-(c^\dagger_{\bf r}
\mbox{\boldmath$\sigma$} c_{\bf r})^2/6$ where
  $\mbox{\boldmath$\sigma$}$ denotes the Pauli 
matrices, and use a vector Hubbard-Stratonovich field. Such
decompositions, however, do not reproduce the mean-field results at
the saddle-point level. \cite{Schulz90}  

As noted earlier, \cite{Schulz90,Weng91} this difficulty can be
circumvented by writing $  
n_{{\bf r}\uparrow}   n_{{\bf r}\downarrow} = [(  c^\dagger_{\bf r} 
c_{\bf r})^2 - (  c^\dagger_{\bf  r} \mbox{\boldmath$\sigma$} \cdot
{\bf\Omega}_{\bf r} c_{\bf r})^2]/4$ where ${\bf\Omega_r}$ is an
arbitrary unit vector. Spin-rotation
invariance is maintained by averaging the partition function over all
directions of ${\bf\Omega}_{\bf r}$.  
In a path integral formalism, ${\bf\Omega}_{\bf r}$ becomes a
time-dependent variable. After the Hubbard-Stratonovich
transformation, the partition function is given by $Z=\int
{\cal D}[c^\dagger,c] \int {\cal D}[\Delta_c,\Delta_s,{\bf\Omega}] e^{-S}$
with the action 
\begin{eqnarray}
S &=& S_0 + \sum_{\bf r} \int _0^\beta d\tau \Bigl \lbrace 
\frac{1}{U}\bigl(\Delta_{c{\bf r}}^2+\Delta_{s{\bf r}}^2\bigr)
\nonumber \\ && 
- c^\dagger_{\bf r}\bigl(i\Delta_{c{\bf r}}+\Delta_{s{\bf r}}
\mbox{\boldmath$\sigma$} \cdot {\bf\Omega}_{\bf r} \bigr) c_{\bf r}. 
 \Bigr \rbrace .
\label{action1}
\end{eqnarray}
$S_0$ is the action in the absence of interaction.  
Since charge fluctuations are not critical (even when
$T\to 0$), they can be treated at the saddle point (i.e. Hartree-Fock)
level. Their effect is to renormalize the chemical potential
$\mu$ from $U/2$ to $0$. Eq.~(\ref{action1}) then
corresponds to a spin-fermion model where the fermions interact with
their collective spin degrees of freedom  ($\Delta_{\bf
r}{\bf\Omega}_{\bf r}$). [We now denote $\Delta_{s{\bf r}}$ by
$\Delta_{\bf r}$.] Below the crossover temperature $T_X$, i.e. when
$T\ll T_X$, low-energy excitations correspond to orientational spin
fluctuations described by the unit vector field ${\bf\Omega}_{\bf
r}$. We can then consider $\Delta_{\bf r}$ within a saddle point
approximation, i.e. $\Delta_{\bf r}= \Delta_0(-1)^{\bf r}$, where the
fluctuations of $\Delta_0$ are ignored. In order
to compute the fermion spectral function $A({\bf k},\omega)$, one should
first determine the effective action $S[{\bf\Omega}]$ of the unit
vector field ${\bf\Omega}$.  

\section{Spin fluctuations}
The effective action
$S[{\bf\Omega}]$ is obtained by expanding around the
N\'eel state. We first introduce a new field $\phi$ defined by
$\phi_{\bf r}=R^\dagger_{\bf r} c_{\bf r}$, where $R_{\bf r}$ is a
SU(2)/U(1) matrix which rotates the spin-quantization axis from
$\hat{\bf z}$ to ${\bf\Omega}_{\bf r}$ ($R_{\bf
r}\sigma_zR^\dagger_{\bf r}={\bf\Omega}_{\bf r} \cdot
\mbox{\boldmath$\sigma$}$). In terms of this new field, the action becomes
\begin{eqnarray}
S &=& S_{\rm MF} + \sum_{\bf r} \int_0^\beta d\tau \phi^\dagger_{\bf
  r} R^\dagger_{\bf r} \partial_\tau R_{\bf r} \phi_{\bf r} 
\nonumber \\ &&
-t \sum_{\langle {\bf r},{\bf r}'\rangle} \int_0^\beta d\tau
[\phi^\dagger_{\bf r} (R^\dagger_{\bf r}R_{{\bf r}'}-1)\phi_{{\bf r}'}
  + {\rm c.c.}] ,
\label{action2}
\end{eqnarray}
where $S_{\rm MF}=S_0+\sum_{\bf r}\int d\tau (\Delta_0^2/U-\Delta_0
(-1)^{\bf r}\phi^\dagger_{\bf r}\sigma_z\phi_{\bf r})$. Within a saddle-point
approximation with ${\bf\Omega}_{\bf r}=\hat {\bf z}$ ($R_{\bf r}=1$),
i.e. ignoring spin fluctuations, one 
recovers the mean-field action $S_{\rm MF}$ of 
the N\'eel state. The value of the order parameter,
$\Delta_0=(U/2)(-1)^{\bf r}\langle \phi^\dagger_{\bf r}\sigma_z \phi_{\bf
r}\rangle$, is obtained by minimizing the free energy. In the weak-coupling
limit, this gives $\Delta_0\sim te^{-2\pi\sqrt{t/U}}$.
\cite{Schrieffer89}

Low-energy spin excitations correspond  to fluctuations of the unit
vector field ${\bf\Omega}_{\bf r}$ around its saddle-point value. The
standard procedure\cite{Haldane83,Weng91,Schulz90} is then to assume
at least local AF order and write 
${\bf\Omega}_{\bf r}={\bf n}_{\bf r}(1-{\bf L}^2_{\bf
r})^{1/2}+(-1)^{\bf r}{\bf L}_{\bf r}$, where the (N\'eel) order parameter
field ${\bf n}_{\bf r}$ is slowly varying in space and time and ${\bf
L}_{\bf r}$ is a small canting field ($|{\bf n}_{\bf r}|=1$, 
${\bf L}_{\bf r}\cdot {\bf
n}_{\bf r}=0$ and $|{\bf L}_{\bf r}|\ll 1$). Integrating out both $\phi$ and
${\bf L}$ yields the action of the NL$\sigma$M.
\cite{Schulz90,Borejsza01} In the strong-coupling limit $U\gg t$, one
recovers the action derived from the Heisenberg model. 

As we now verify explicitly, the small canting field ${\bf L}_{\bf r}$
gives negligible contributions to the parameters of the NL$\sigma$M
in the weak-coupling limit $U\ll t$. If we identify ${\bf\Omega}_{\bf
r}$ with the slowly varying N\'eel field, ${\bf\Omega}_{\bf
r}\approx {\bf n}_{\bf r}$, the effective action $S[{\bf n}]$ is
readily obtained. 
Integrating out the fermions in Eq.~(\ref{action2}) and taking the
continuum limit in space, one obtains to
lowest-order in gradient (i.e. in $\partial_\tau R$ and
$\mbox{\boldmath$\nabla$}_{\bf r}R$) \cite{Dupuis00}  
\begin{equation}
S[{\bf n}] = \frac{1}{2} \int d^2r d\tau \bigl[ \chi^0_\perp
(\partial_\tau {\bf n})^2+ \rho_s^0 (\mbox{\boldmath$\nabla$}_{\bf r}
{\bf n})^2 \bigr],
\label{action3}
\end{equation}
where $\chi^0_\perp$ is the uniform transverse spin susceptibility in
the mean-field state and $\rho_s^0=-(\langle
K\rangle_{\rm MF}/2+\Pi^0_\perp)/4$ the spin stiffness. Here $\langle
K\rangle_{\rm MF}$
is the mean-value of the kinetic energy and $\Pi^0_\perp$ the
correlation function of the transverse spin current ($j^x$ or
$j^y$). Eq.~(\ref{action3}) should be supplemented with a
short-distance cutoff (in momentum space) $\Lambda\sim \xi^{-1}_0$, since
short-range AF order cannot be defined at length scales smaller that
the coherence length $\xi_0\sim t/\Delta_0$. Using the mean-field
action $S_{\rm MF}$, one obtains ($N$ is the number of lattice sites)
\begin{eqnarray}
\chi^0_\perp &=& \frac{\Delta_0^2}{4N} \sum_{\bf k} \frac{1}{E^3_{\bf
k}} \sim \frac{1}{t} \sqrt{\frac{t}{U}}  , \label{chi0} \\ 
\rho^0_s &=& \frac{t^2\Delta_0^2}{N} \sum_{\bf k} \frac{\sin
^2k_x}{E^3_{\bf k}} \sim t, \label{rho0} 
\end{eqnarray}
where $E_{\bf k}=(\epsilon^2_{\bf k}+\Delta_0^2)^{1/2}$ is the
Bogoliubov quasi-particle excitation energy in the mean-field state
($\epsilon_{\bf k}=-2t(\cos k_x+\cos k_y)$ is the dispersion of the
free fermions). We can verify that Eqs.~(\ref{action3}-\ref{rho0}) can
be directly obtained from the results of
Refs.~\cite{Schulz90,Borejsza01} in the weak-coupling limit ($U\ll t$). 
The value of the spin-wave velocity $c=\sqrt{\rho^0_s/\chi^0_\perp}\sim
t(U/t)^{1/4}$ also agrees with the weak-coupling limit of the RPA result.
\cite{Chubukov92} The approximation ${\bf\Omega}_{\bf r}\approx {\bf
n}_{\bf r}$ is
therefore justified when $U\ll t$. While it restricts the validity of
our approach to the weak-coupling limit, it makes the computation of
fermionic correlation functions considerably simpler, since the
fermions couple directly to the N\'eel field [see Eq.~(\ref{action1})].

We solve the NL$\sigma$M within a ``large-$\cal N$'' approach
by extending the number of components of the unit vector
${\bf n}_{\bf r}$ from 3 to $\cal N$. When $\cal N\to \infty$, the
action (\ref{action3}) can be solved exactly by a
saddle-point method. \cite{Sachdev} Fig.~1 shows the resulting
crossover diagram 
as a function of the dimensionless coupling constant $\bar g=\Lambda
g= \Lambda c{\cal N}/\rho^0_s$ of the NL$\sigma$M. In the
weak-coupling limit of the Hubbard model ($U\ll t$), $\bar
g=c\Delta_0{\cal N}/(\rho^0_st) \propto 
e^{-2\pi\sqrt{t/U}}$ is exponentially small. This implies that the
ground state has AF long-range order with very weak quantum
fluctuations. This magnetic order persists in the strong-coupling
regime ($U\gg t$) where\cite{Schulz90} $\bar g\stackrel{\textstyle
<}{\sim} \bar g_c=4\pi$  (see Fig.~1) in agreement with conclusions 
based on the Heisenberg model (for a square lattice). At finite
temperature, magnetic 
long-range order is suppressed as required by the Mermin-Wagner
theorem. The dominant fluctuations are classical since the gap $m$ in
the spin excitation spectrum (see below) is much smaller than the temperature
(this regime is known as ``renormalized classical'' in the literature
\cite{Chakravarty89}).   

Since we are primarily interested in the fermion spectral function $A({\bf
k},\omega)$ at finite temperature, we shall consider the action
$S[{\bf n}]$ in this regime. In the large-$\cal N$ limit, it reads
\begin{equation}
S[{\bf n}] =  \frac{\cal N}{2gc} \sum_{{\bf q},\omega_\nu} 
(\omega_\nu^2+c^2q^2+m^2) |{\bf n}({\bf q},i\omega_\nu)|^2 ,
\label{action4}
\end{equation}
where we have introduced the Fourier transformed field ${\bf n}({\bf
q},i\omega_\nu)$ ($\omega_\nu$ is a bosonic Matsubara frequency). The
length of the vector ${\bf n}_{\bf r}$ is no 
longer fixed to unity. In the large-$\cal N$ solution, the constraint
$|{\bf n}_{\bf r}|=1$ is imposed only on average (via the Lagrange multiplier
$m$).\cite{Sachdev} The mass $m$ of the spin fluctuation propagator
($\alpha=1\cdots \cal N$)  
\begin{eqnarray}
\chi({\bf q},i\omega_\nu) &=& \langle {\bf n}_\alpha({\bf q},i\omega_\nu)
  {\bf n}_\alpha(-{\bf q},-i\omega_\nu) \rangle \nonumber \\ 
&=& \frac{gc/{\cal N}}{\omega_\nu^2+c^2q^2+m^2} 
\label{chi}
\end{eqnarray}
is determined by the saddle-point equation
\begin{equation}
1=gc\frac{T}{N}\sum_{{\bf q},\omega_\nu} 
\frac{1}{\omega_\nu^2+c^2q^2+m^2}.
\label{SP}
\end{equation}
In the renormalized classical regime, we can neglect quantum
fluctuations. This approximation is excellent in the weak-coupling
regime ($U\ll t$) since quantum fluctuations are weak ($\bar g\ll \bar
g_c$, see Fig.~1). From Eq.~(\ref{SP}), we then obtain the AF coherence
length $\xi=c/m\sim \Lambda^{-1} \exp(2\pi\rho^0_s/{\cal N}T)$. 

Note that we expect also a term $m^2|\omega_\nu|/\omega_{\rm sf}$ in
the denominator in Eq.~(\ref{chi}). This term comes from the damping
of spin fluctuations by gapless fermion excitations. \cite{Millis90} It is
missed in our approach since we expand around the zero-temperature AF
state which has only 
gapped quasi-particle excitations. Fluctuations are classical when
$m\ll T$ and $\omega_{\rm sf}\ll T$. Both conditions are satisfied
in the renormalized classical regime ($T\ll T_X$) since $\omega_{\rm
sf}\sim \xi^{-2}\to 0$ (critical slowing down).
\cite{Vilk97,Vilk97a,Millis90}

\section{Spectral function}

 Knowing the effective action $S[{\bf n}]$ of
the spin excitations [Eq.~(\ref{action4})], we are now in a position
to compute the spectral function $A({\bf k},\omega)=-\pi^{-1}{\rm
Im}G({\bf k},\omega)$ from the spin-fermion model
(\ref{action1}). Here $G({\bf k},\omega)$ denotes the retarded part of
the fermionic Green's function. By integrating first the fermions and
then the spin fluctuations, we can write the Green's function as
\begin{equation}
G({\bf r}-{\bf r}',\tau-\tau') = \frac{1}{Z} \int {\cal D}[{\bf n}] 
e^{-S[{\bf n}]} G({\bf r},\tau;{\bf r}',\tau'|{\bf n}) .
\label{GF}
\end{equation}
$G({\bf r},\tau;{\bf r}',\tau'|{\bf n})$ is the Green's function
for a given configuration of ${\bf n}$: $G^{-1}[{\bf n}]=
G_0^{-1}+\Delta_0(-1)^{\bf r} \mbox{\boldmath$\sigma$}\cdot{\bf n}_{\bf r}$,
where $G_0$ is the Green's function of the free fermions. Since
$S[{\bf n}]$ is Gaussian in the large-$\cal N$ limit, the averaging in
Eq.~(\ref{GF}) is
easily done. The result can be written as $G^{-1}({\bf k},i\omega_n)
=G_0^{-1}({\bf k},i\omega_n)-\Sigma({\bf k},i\omega_n)$ ($\omega_n$ is
a fermionic Matsubara frequency). 

We consider the lowest-order contribution to the self energy $\Sigma$
(Fig.~2):
\begin{eqnarray}
\Sigma({\bf k},i\omega_n) &=& \Delta^2_0 \frac{T}{N} \sum_{{\bf
q},\omega_\nu} {\cal N} \chi({\bf q},i\omega_\nu)
G_0({\bf k}-{\bf Q}-{\bf q},i\omega_n-i\omega_\nu) \nonumber \\ 
& \simeq & \Delta_0^2 \frac{gT}{cN} \sum_{\bf q}
\frac{1}{q^2+\xi^{-2}}  \frac{1}{i\omega_n-\epsilon_{{\bf k}-{\bf
      Q}-{\bf q}}} ,
\label{Sig1}
\end{eqnarray}
where the last line has been obtained in the classical limit
($\omega_\nu=0$) and ${\bf Q}=(\pi,\pi)$. At low temperature when
$\xi\to\infty$, the sum over 
${\bf q}$ in Eq.~(\ref{Sig1}) diverges in 2D due to the contribution of
long wavelengths (${\bf q}\sim 0$). We can therefore expand
$-\epsilon_{{\bf k}-{\bf Q}-{\bf q}}=\epsilon_{{\bf k}-{\bf
q}}\simeq \epsilon_{\bf k}-{\bf v}_{\bf k}\cdot {\bf q}$ around ${\bf
q}=0$ (${\bf v}_{\bf k}$ is the velocity of the free
fermions). Let us first consider a particle at the Fermi 
level. One easily finds that the imaginary part of the retarded
self-energy ($i\omega_n\to \omega+i0^+$) takes the form
\begin{equation}
\Sigma''({\bf k}_F,\omega=0) \approx -
\frac{\Delta_0^2\xi}{\rho_s^0\xi_{\rm th}} \propto -T\xi ,
\end{equation} 
where $\xi_{\rm th}=|{\bf v}_{\bf k}|/T$ is the De Broglie thermal
wavelength. Since $\xi$ grows exponentially below $T_X$, it quickly
becomes larger than $\xi_{\rm th}$. As a result, 
$\lim_{T\to 0} \xi/\xi_{\rm th}=\infty$ and $\Sigma''({\bf
k}_F,\omega=0)$ diverges at low temperature in contradiction with
the Fermi-liquid theory hypothesis. Thus, the lowest-order perturbation
result shows that quasi-particles are suppressed by spin fluctuations
when $T\ll T_X$. This phenomenon is accompanied by the formation of a
pseudogap. For $|\omega+\epsilon_{\bf k}|\gg |{\bf v}_{\bf k}|/\xi$, the
real and imaginary parts of the self-energy are given by\cite{note3}
\begin{equation}
\Sigma'({\bf k},\omega)\simeq \frac{\Delta_0^2}{\omega+\epsilon_{\bf k}} ,
\,\,\,\,   \Sigma''({\bf k},\omega)\simeq -\frac{3\Delta_0^2T}{4\pi
  \rho_s^0|\omega+\epsilon_{\bf k}|} .
\label{Sig2}
\end{equation}
Note that the condition $|\omega+\epsilon_{\bf k}|\gg |{\bf v}_{\bf
  k}|/\xi$ is satisfied for any value of $\omega$ except in an
  exponentially small window around $\omega=-\epsilon_{\bf k}$. From
  Eq.~(\ref{Sig2}), we deduce the spectral function 
\begin{equation}
A({\bf k},\omega) = \frac{\gamma}{\pi} \frac{|\omega+\epsilon_{\bf
    k}|}{(\omega^2-E_{\bf k}^2)^2+\gamma^2} , \,\,\,\, \gamma \approx
    \frac{3\Delta_0^2T}{4\pi\rho_s^0} . 
\label{sf}
\end{equation}
$A({\bf k},\omega)$ exhibits two peaks at $\pm E_{\bf k}$ that are
precursors of the AF bands that exist in the $T=0$ ordered state. The
width of these peaks is given by $\gamma/\Delta_0\sim
T\Delta_0/\rho_s^0\sim Te^{-2\pi\sqrt{t/U}}$. The precursors of the AF
bands are separated by a pseudogap. In particular $A({\bf
k}_F,\omega)$ vanishes at $\omega=0$. 

When $T\to 0$ ($\gamma\to 0$),
\begin{equation}
A({\bf k},\omega) \to \frac{1}{2}
\Bigl(1+\frac{\epsilon_{\bf k}}{E_{\bf k}}\Bigr)\delta(\omega-E_{\bf k}) 
+\frac{1}{2}
\Bigl(1-\frac{\epsilon_{\bf k}}{E_{\bf k}}\Bigr)\delta(\omega+E_{\bf k}), 
\label{sf1}
\end{equation}
which is the spectral function of the $T=0$ AF state. Thus, the simple
self-energy (\ref{Sig1}) predicts that the pseudogap evolves smoothly
into the gap of the ground-state when $T\to 0$. It should be noted that
neglecting quantum fluctuations is justified only at low energy
$|\omega|<T$. In particular, the precise location of the peaks around
$\pm\Delta_0$ should depend on quantum fluctuations since 
$\Delta_0\sim T_X\gg T$. 

The spectral function $A({\bf k},\omega)$ [Eq.~(\ref{sf})] is similar
to the result of the TPSC theory. \cite{Vilk96,Vilk97,Vilk97a} In the latter,
the position of the maxima in $A({\bf k}_F,\omega)$ scales with the
zero-temperature gap, \cite{Dare96,note4} and the width of these two peaks is
proportional to $T$. \cite{Vilk96,Vilk97,Vilk97a} These two features agree
with our conclusions. This similarity is not surprising since in both
approaches a  
paramagnon-like self-energy [Eq.~(\ref{Sig1})] with a similar spin
susceptibility [Eq.~(\ref{chi})] is used to obtain the spectral
function. The main difference comes from the spin fluctuation
propagator $\chi$. While $\chi$ comes from the
NL$\sigma$M (which is itself based on an expansion around the ordered
AF state), it is obtained by considering the paramagnetic phase in the
TPSC theory. As a result, the basic parameters entering the spectral function
$A({\bf k},\omega)$ [Eq.~(\ref{sf})], namely the $T=0$ order parameter
$\Delta_0$ and the $T=0$ spin stiffness $\rho^0_s$, do not appear in
the TPSC theory. Instead, $A({\bf k},\omega)$ is expressed only in terms of
the paramagnetic properties of the system. 

Two comments are in order here. The validity of Eq.~(\ref{Sig1}),
which does not include vertex correction, may be
questioned. \cite{note5} The
importance of these corrections is a long-standing problem which is
still under debate. Vertex corrections are expected to play a crucial
role when higher-order self-energy contributions are taken into
account. The FLEX approximation, which sums up contributions to all
order without vertex correction, does not predict the formation of a
pseudogap in $A({\bf k},\omega)$ at low temperature \cite{Bickers89}
(see Ref.~\onlinecite{Vilk97} for a detailed discussion of the FLEX
approximation).  

In the spin-fermion model defined by Eqs.~(\ref{action1}) and
(\ref{action3}), there are only two (transverse) spin excitation
modes, as expected when only orientational fluctuations are important
($T\ll T_X$). Unfortunately, this property is lost in the large-$\cal 
N$ limit of the NL$\sigma$M [Eq.~(\ref{action4})], where both
transverse and amplitude fluctuations are allowed. Following
Ref.~\onlinecite{Schmalian99}, $A({\bf k},\omega)$ can be obtained exactly
when $\xi\to\infty$ by summing all the self-energy diagrams. The result,
\begin{eqnarray}
A({\bf k},\omega) &=& \frac{3^{\frac{3}{2}}}{\sqrt{2\pi}\Delta_0^3}
(\omega^2-\epsilon_{\bf k}^2)^{1/2}(\omega+\epsilon_{\bf k})
\exp \Bigl( -\frac{3}{2}\frac{\omega^2-\epsilon^2_{\bf k}}{\Delta_0^2}
\Bigr) 
\nonumber \\ && \times
\bigl[ \theta(\omega-|\epsilon_{\bf k}|) -
 \theta(-\omega-|\epsilon_{\bf k}|) \bigr] ,
\end{eqnarray}
shows two broad incoherent features, located around
$\pm\sqrt{2/3}\Delta_0$ for $\epsilon_{\bf k}=0$, instead
of the correct $T=0$ limit given by Eq.~(\ref{sf1}). The correct
limit is obtained only when amplitude fluctuations are frozen in the limit
$\xi\to\infty$. \cite{Tchernyshyov99,Monien01} We therefore conclude that our
approach, which is based on the large-$\cal N$ solution of the
NL$\sigma$M, must break down at very low temperature. The fact that
the spectral function $A({\bf k},\omega)$ derived from the
lowest-order self-energy contribution does reproduce the correct
result when $T\to 0$ [Eq.~(\ref{sf})] appears somewhat accidental. A correct
treatment of the $T\to 0$ limit must freeze the amplitude
fluctuations of the N\'eel field ${\bf n}$.  

\section{Conclusion}

We have described a new approach to the pseudogap in the half-filled
2D Hubbard model at weak coupling. Within this approach, only
orientational spin fluctuations are considered, whereas fluctuations
of the amplitude of the local spin density are ignored. This
approximation is justified below a crossover temperature $T_X$ (of the
order of the mean-field AF transition temperature) where the AF
correlation length starts to grow exponentially (renormalized
classical regime). The effective action of spin fluctuations is then
given by a NL$\sigma$M. Solving the NL$\sigma$M within a ``large-$\cal
N$'' approach, we find that the ground-state of the Hubbard model on a
square lattice is antiferromagnetic (N\'eel order) for any value of
the Coulomb interaction $U$ (Fig.~1). \cite{Schulz90} 

We have obtained the fermion spectral function $A({\bf k},\omega)$ in
the weak-coupling limit by computing the self-energy $\Sigma({\bf
k},\omega)$ to lowest order in the spin-fermion interaction
(Fig.~2). The QP peak which characterizes the Fermi liquid state is
suppressed by spin fluctuation when $T\ll T_X$. Instead, $A({\bf
  k},\omega)$ exhibits a pseudogap separating two broadened
peaks. These peaks are precursors of the Bogoliubov QP's that appear
at the $T=0$ AF transition. Our results are in very good
agreement with those obtained by the TPSC theory.
\cite{Vilk96,Vilk97,Vilk97a}  An important limitation of
our analysis comes from the large-$\cal N$ solution of the
NL$\sigma$M. The latter introduces amplitude fluctuations of the
N\'eel field which should be frozen at low temperature. As a result,
when going beyond the lowest-order contribution to $\Sigma({\bf
k},\omega)$, we do not obtain the correct $T\to 0$ limit of the
fermion spectral function. In Ref.~\onlinecite{Borejsza01}, we show
how this difficulty can be circumvented. 

There are several directions in which this work could be further
developed. Since the NL$\sigma$M description is valid both at weak
($U \ll t$) and strong ($U\gg t$) coupling, our analysis of the
fermion spectral function could be extended in the regime $U\gg t$. In
the Mott-Hubbard insulator, we expect the pseudogap to transform into
a (charge) gap of order $U$, the precursors of the $T=0$ AF bands
becoming the upper and lower Hubbard bands.

It is also possible to consider variants of the square lattice Hubbard
model [Eq.~\ref{Ham}] where antiferromagnetism becomes frustrated. This
would be the case for the $t-t'$ Hubbard model ($t'$ is the hopping
amplitude for next-nearest neighbors) or if the lattice is triangular
instead of square. Doping may also induce some kind of magnetic
frustration. \cite{note6} This opens up the possibility to reach the quantum
disordered and quantum critical regimes of the NL$\sigma$M (Fig.~1)
and to study the corresponding fermion spectral functions.

\acknowledgments

I would like to thank A.-M. Tremblay for useful discussions and a
critical reading of the manuscript.

\begin{figure}
\epsfxsize 8cm
\epsffile[40 330 340 530]{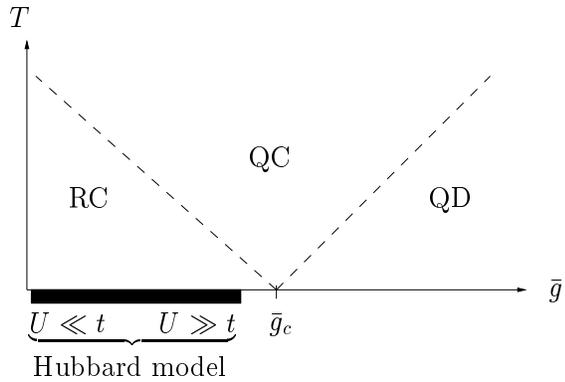}
\caption{Crossover diagram derived from the large-$\cal N$ limit of the 2D 
NL$\sigma$M. At $T=0$, there is long-range order when  the
dimensionless coupling $\bar g\leq \bar g_c=4\pi$. The three
finite-temperature regimes correspond to ``renormalized classical''
(RC), ``quantum critical'' (QC) and ``quantum disordered'' (QD).
\cite{Chakravarty89} The ground-state of the half-filled 2D Hubbard
model on a square lattice is
ordered for any value of the Coulomb repulsion $U$. At finite
temperature, there are strong AF fluctuations with 
an exponentially large coherence length (RC regime). }
\label{Fig1}
\end{figure}

\begin{figure}
\epsffile[220 360 400 440]{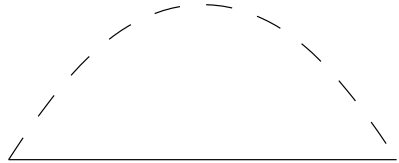}
\caption{Lowest-order contribution to the fermion self-energy $\Sigma$. The
dashed line represents the spin propagator $\chi$ [Eq.~(\ref{chi})]. }
\label{Fig2} 
\end{figure}

\ecols 


\begin{thebibliography}{0}
\bibitem{note1} Note that the pseudogap issue resurfaces in the
  strong-coupling limit when the system is doped away from
  half-filling. It is generally believed that the spectral function of
  the doped holes exhibits a pseudogap governed by the energy scale $J$.
\bibitem{note5} Vertex
corrections are included in the TPSC theory: they do not lead to any
qualitative change. \cite{Vilk96,Vilk97,Vilk97a}
\bibitem{note6} Doping may not be easily included in our
  approach. When deriving the NL$\sigma$M, one should first solve the
  Hartree-Fock theory. Away from half-filling, the Hartree-Fock
  ground-state is not definitely known. Many suggestions can be found
  in the literature: incommensurate spin-density wave, spiral phase,
  stripes, etc.
\end{thebibliography}
\end{document}